\documentclass[12pt]{article}

\usepackage{graphicx}
\begin{document}

\begin{center}
{\bf 4D Einstein--Gauss--Bonnet gravity coupled to modified logarithmic nonlinear electrodynamics} \\
\vspace{5mm} Sergey Il'ich Kruglov
\footnote{E-mail: serguei.krouglov@utoronto.ca}
\underline{}
\vspace{3mm}

\textit{Department of Physics, University of Toronto, \\60 St. Georges St.,
Toronto, ON M5S 1A7, Canada\\
Department of Chemical and Physical Sciences, University of Toronto,\\
3359 Mississauga Road North, Mississauga, Ontario L5L 1C6, Canada}

\vspace{5mm}
\end{center}
\begin{abstract}
Spherically symmetric solution in 4D Einstein--Gauss--Bonnet gravity coupled to modified logarithmic nonlinear electrodynamics (ModLogNED) is found. This solution at infinity possesses the charged black hole Reissner--Nordstr\"{o}m behavior. We study the black hole thermodynamics, entropy, shadow, energy emission rate and quasinormal modes. It was shown that black holes can possess the phase transitions and at some range of event horizon radii black holes are stable. The entropy has the logarithmic correction to the area law. The shadow radii were calculated for variety of parameters. We found that there is a peak of the black hole energy emission rate. The real and imaginary parts of the quasinormal modes frequencies were calculated. The energy conditions of ModLogNED are investigated.
\end{abstract}

\vspace{3mm}
\textit{Keywords}: Einstein$-$Gauss$-$Bonnet gravity; nonlinear electrodynamics; Hawking temperature; entropy; heat capacity; black hole shadow;
energy emission rate; quasinormal modes
\vspace{3mm}

\section{Introduction}

Nowadays, there are many theories of gravity that are alternatives to Einstein’s theory \cite{Clifton, Will}. The motivation of generalisations of Einstein's theory of General Relativity (GR) is to resolve some problems in cosmology and astrophysics. One of important modification of GR is the Einstein--Gauss--Bonnet (EGB) theory \cite{Lanczos,Lanczos1,Lovelock,Lovelock1}. EGB theories do not include extra degrees of freedom and field equations have second derivatives of the metric. These theories also prevent Ostrogradsky instability \cite{Ostrogradsky}.
The four dimensional (4D) EGB theory, that includes the Einstein--Hilbert action plus GB term, is a particular case of the Lovelock theory. It represents the generalization of Einstein's GR for higher dimensions and EGB theory results covariant second-order field equations. The GB part of the action possesses higher order curvature terms. It is worth mentioning that at low energy the action of the heterotic string theory includes higher order curvature terms \cite{Witten,Gross,Metsaev,Metsaev1,Zwiebach}. Therefore, it is of interest to study gravity action with the GB term. The GB term is a topological invariant in 4D and before a regularization it does not contribute to the equation of motion. But
Glavan and Lin \cite{Glavan} showed  that re-scaling the coupling constant, after the regularization, GB term contributes to the equation of motion.
The consistent theory of 4D EGB gravity, was proposed in \cite{Aoki,Aoki1,Aoki2}, is in agreement with the Lovelock theorem \cite{Lovelock} and possesses two dynamical degrees of freedom  breaking the temporal diffeomorphism invariance.
It is worth noting that the theory of \cite{Aoki,Aoki1,Aoki2}, in the spherically-symmetric metrics, gives the solution which is a solution in the framework of \cite{Glavan} scheme (see \cite{Lobo1}). Some aspects of 4D EGB gravity were considered in \cite{ Fernandes2}.
The black hole and wormhole type solutions in the effective gravity models, including higher curvature
terms, were obtained in \cite{Alexeyev}.

Here, we study the black hole thermodynamics, the entropy, the shadow, the energy emission rate and quasinormal modes in the framework of the
ModLogNED model (proposed in \cite{Krug2}) coupled to 4D EGB gravity. It is worth noting that ModLogNED model is simpler compared with logarithmic model \cite{Soleng} and generalized logarithmic model \cite{Kruglov5} because the mass and metric functions here are expressed through simple elementary functions.
The black hole quasinormal modes, deflection angles, shadows  and the Hawking radiation were studied in \cite{Konoplya1,Konoplya2,Belhaj1,Konoplya3,Stefanov,Guo1,Wei1}.

The structure of the paper is as follows. In Sect. 2, we obtain the spherically symmetric solution of black holes in the 4D EGB gravity coupled to ModLogNED. At infinity the Reissner$-$Nordstr\"{o}m behavior of the charged black holes takes place. The black hole thermodynamics is studied in Sect. 3. We calculate the Hawking temperature, the heat capacity and the entropy. At some parameters second order phase transitions occur. The entropy includes the logarithmic correction to Bekenstein--Hawking entropy. In Sect. 4 the black hole shadow is investigated. We calculate the photon sphere, the event horizon, and the shadow radii. The black hole energy emission rate is  investigate in Sect. 5.  In Sect. 6 we study quasinormal modes and find complex frequencies. Section 7 is a summary. In Appendix A energy conditions of ModLogNED model are investigated.

\section{4D EGB model}

The action of EGB gravity coupled to nonlinear electrodynamics (NED) in D-dimensions is given by
\begin{equation}
I=\int d^Dx\sqrt{-g}\left[\frac{1}{16\pi G}\left(R+ \alpha{\cal L}_{GB}\right)+{\cal L}_{NED}\right],
\label{1}
\end{equation}
where $G$ is the Newton's constant, $\alpha$ has the dimension of (length)$^2$. The Lagrangian of ModLogNED, proposed in \cite{Krug2}, is
\begin{equation}
{\cal L}_{NED} = -\frac{\sqrt{2{\cal F}}}{8\pi\beta}\ln\left(1+\beta\sqrt{2{\cal F}}\right),
 \label{2}
\end{equation}
where we use Gaussian units.
The parameter $\beta$ ($\beta\geq 0$) possesses the dimension of (length)$^2$, $F_{\mu\nu}=\partial_\mu A_\nu-\partial_\nu A_\mu$ is the field strength tensor, and ${\cal F}=(1/4)F_{\mu\nu}F^{\mu\nu}=(B^2-E^2)/2$, where $B$ and $E$ are the induction magnetic and electric fields, correspondingly. Making use of the limit $\beta\rightarrow 0$ in Eq. (2), we arrive at the Maxwell's Lagrangian ${\cal L}_M = -{\cal F}/(4\pi)$. The GB Lagrangian has the structure
\begin{equation}
{\cal L}_{GB}=R^{\mu\nu\alpha\beta}R_{\mu\nu\alpha\beta}-4R^{\mu\nu}R_{\mu\nu}+R^2.
\label{3}
\end{equation}
By varying action (1) with respect to the metric we have EGB equations
\begin{equation}
R_{\mu\nu}-\frac{1}{2}g_{\mu\nu}R+\alpha H_{\mu\nu}=-8\pi GT_{\mu\nu},
\label{4}
\end{equation}
\begin{equation}
H_{\mu\nu}=2\left(RR_{\mu\nu}-2R_{\mu\alpha}R^\alpha_{~\nu}-2R_{\mu\alpha\nu\beta}R^{\alpha\beta}-
R_{\mu\alpha\beta\gamma}R^{\alpha\beta\gamma}_{~~~\nu}\right)-\frac{1}{2}{\cal L}_{GB}g_{\mu\nu},
\label{5}
\end{equation}
where $T_{\mu\nu}$ is the stress (energy-momentum) tensor. To obtain the solution of field equations we need to use an ansatz for the interval. But the validity of Birkhoff’s theorem \cite{Birkhoff} for our case of 4D EGB gravity coupled to ModLogNED model is not proven. Therefore, to simplify the problem we consider magnetic black holes with the static spherically symmetric metric in $D$ dimension. In addition, we assume that components of the interval are restricted by the relation $g_{11}=g_{00}^{-1}$. Thus, we suppose that the metric has the form
\begin{equation}
ds^2=-f(r)dt^2+\frac{dr^2}{f(r)}+r^2d\Omega^2_{D-2}.
\label{6}
\end{equation}
The $d\Omega^2_{D-2}$ is the line element of the unit $(D - 2)$-dimensional sphere. By following
\cite{Glavan} we replace $\alpha$ by $\alpha\rightarrow \alpha/(D-4)$ and taking the limit $D\rightarrow 4$.
We study the magnetic black holes and find ${\cal F}=q^2/(2r^4)$, where $q$ is a magnetic charge. Then the magnetic energy density becomes \cite{Krug2}
\begin{equation}\label{7}
  \rho=T_0^{~0}=-{\cal L} =\frac{\sqrt{2{\cal F}}}{8\pi\beta}\ln\left(1+\beta\sqrt{2{\cal F}}\right)=\frac{q}{8\pi\beta r^2}\ln\left(1+\frac{\beta q}{r^2}\right).
\end{equation}
At the limit $D \rightarrow 4$ and from Eq. (4) we obtain
\begin{equation}
r(2\alpha f(r)-r^2-2\alpha)f'(r)-(r^2+\alpha f(r)-2\alpha)f(r)+r^2-\alpha=2r^4G\rho.
\label{8}
\end{equation}
By virtue of Eq. (7 ) one finds
\begin{equation}\label{9}
4\pi\int_0^r r^2\rho dr=m_M+\frac{q}{2\beta}\left[r\ln\left(1+\frac{\beta q}{r^2}\right)-2\sqrt{\beta q}\arctan\left(\frac{\sqrt{\beta q}}{r}\right)\right],
\end{equation}
\begin{equation}\label{10}
m_M=4\pi\int_0^\infty r^2\rho dr=\frac{q}{2\beta}\int_0^\infty \ln\left(1+\frac{\beta q}{r^2}\right)dr=\frac{\pi q^{3/2}}{2\sqrt{\beta}},
\end{equation}
where $m_M$ is the black hole magnetic mass.
Making use of Eqs. (9) and (10) we obtain the solution to Eq. (8)
\[
f(r)=1+\frac{r^2}{2\alpha}\left(1\pm\sqrt{1+\frac{8\alpha G}{r^3}(m+h(r)}\right),
\]
\begin{equation}
h(r)=m_M+\frac{q}{2\beta}\left[r\ln\left(1+\frac{\beta q}{r^2}\right)-2\sqrt{\beta q}\arctan\left(\frac{\sqrt{\beta q}}{r}\right)\right],
\label{11}
\end{equation}
where $m$ is the constant of integration (the Schwarzschild mass) and the total black hole mass is $M=m+m_M$ which is the ADM mass.
At the limit $\beta\rightarrow 0$ one has
\[
\lim_{\beta\rightarrow 0}h(r)=m_M-q^2/2r.
\]
Then making use of Eq. (11), for the negative branch, we obtain
\[
\lim_{\beta\rightarrow 0,\alpha\rightarrow 0}f(r)=1-\frac{2MG}{r}+\frac{Gq^2}{r^2},
\]
that corresponds to GR coupled to Maxwell electrodynamics (the Reissner--Nordstr\"{o}m solution).

It is worth mentioning that for spherically symmetric $D$-dimensional line element (6), the Weyl tensor of the $D$-dimensional spatial part becomes zero \cite{Lobo1}. Therefore, solution (11) corresponds to the consistent theory \cite{Aoki,Aoki1,Aoki2}.
By introducing  the dimensionless variable $x= r/\sqrt{\beta q}$, Eq. (11) is rewritten in the form
\begin{equation}
f(x)=1+Cx^2\pm C\sqrt{x^4+x(A-Bg(x))},
\label{12}
\end{equation}
where
\begin{equation}\label{13}
A=\frac{8\alpha GM}{(\beta q)^{3/2}},~B=\frac{4\alpha G}{\beta^2},~C=\frac{\beta q}{2\alpha},~ g(x)=2\arctan\left(\frac{1}{x}\right)-x\ln\left(1+\frac{1}{x^2}\right).
\end{equation}
We will use the negative branch in Eqs. (11) and (12) with the minus sign of the square root to have black holes without ghosts.
As $\alpha\rightarrow 0$, $r\rightarrow \infty$ the metric function $f(r)$ (11), for the negative branch, becomes
\begin{equation}
f(r)=1-\frac{2MG}{r}+\frac{Gq^2}{r^2}+{\cal O}(r^{-3}),
\label{14}
\end{equation}
showing, at infinity, the Reissner$-$Nordstr\"{o}m behavior of the charged black holes. The plot of function (12) for a particular chose of parameters, $A=15$, $C=1$ (as an example), is depicted in Fig. 1.
\begin{figure}[h]
\includegraphics[height=4.0in,width=4.0in]{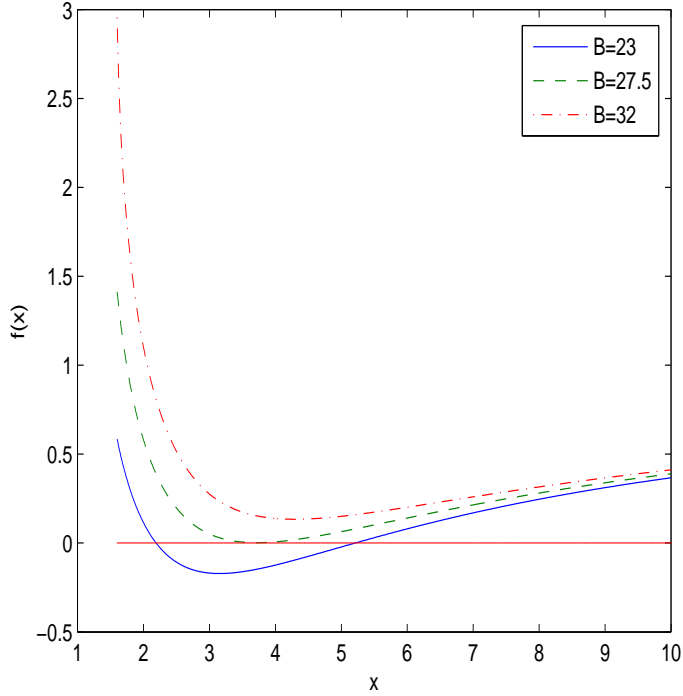}
\caption{\label{fig.1}The plot of the function $f(x)$ for $A=15, C=1$.}
\end{figure}
The expansion (14) was observed in other models (see, for example, \cite{Mignemi}). According to Fig. 1 there can be two horizons or one (the extreme) horizon of black holes.

\section{The black hole thermodynamics}

To study the black hole thermal stability we will calculate the Hawking temperature
\begin{equation}
T_H(r_+)=\frac{f'(r)\mid_{r=r_+}}{4\pi},
\label{15}
\end{equation}
where $r_+$ is the event horizon radius ($f(r_+)=0$). From Eqs. (12) and (15) one finds the Hawking temperature
\begin{equation}
T_H(x_+)=\frac{1}{4\pi \sqrt{\beta q}}\left(\frac{2Cx_+^2-1+BC^2x_+^2g'(x_+)}{2x_+(1+Cx_+^2)}\right),
\label{16}
\end{equation}
\[
g'(x_+)=-\ln\left(1+\frac{1}{x_+^2}\right).
\]
Parameter $A$ was substituted into Eq. (15) from equation $f(x_+)=0$.
The plot of the dimensionless function $T_H(x_+)\sqrt{\beta q}$ versus $x_+$, for the case $C=1$, is represented in Fig. 2.
\begin{figure}[h]
\includegraphics[height=4.0in,width=4.0in]{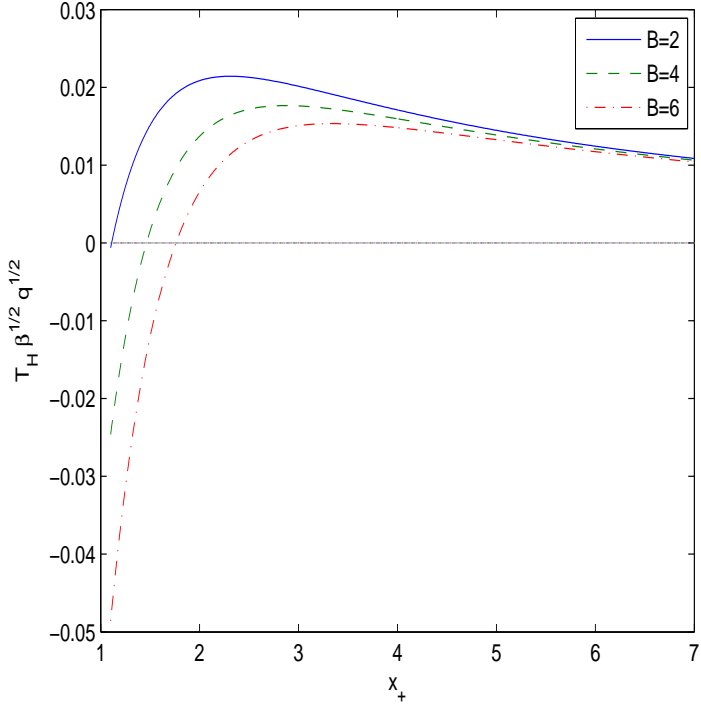}
\caption{\label{fig.2}The plot of the function $T_H(x_+)\sqrt{\beta q}$ at $C=1$.}
\end{figure}
Figure 2 shows that the Hawking temperature is positive for some interval of event horizon radii.
 We will calculate the heat capacity to study the black hole local stability
\begin{equation}
C_q(x_+)=T_H\left(\frac{\partial S}{\partial T_H}\right)_q=\frac{\partial M(x_+)}{\partial T_H(x_+)}=\frac{\partial M(x_+)/\partial x_+}{\partial T_H(x_+)/\partial x_+},
\label{17}
\end{equation}
where $M(x_+)$ is the black hole gravitational mass as a function of the event horizon radius. Making use of equation $f(x_+)=0$ we obtain the black hole mass
\begin{equation}
M(x_+)=\frac{(\beta q)^{3/2}}{8\alpha G}\left(\frac{1+2Cx_+^2}{C^2x_+}+Bg(x_+)\right).
 \label{18}
\end{equation}
With the help of Eqs. (16) and (18) one finds
\begin{equation}
\frac{\partial M(x_+)}{\partial x_+}=\frac{(\beta q)^{3/2}}{8\alpha G}\left(\frac{2Cx_+^2-1}{C^2x_+^2}+Bg'(x_+)\right),
\label{19}
\end{equation}
\[
\frac{\partial T_H(x_+)}{\partial x_+}=\frac{1}{8\pi \sqrt{\beta q}}\biggl(\frac{5Cx_+^2-2C^2x_+^4+1}{x_+^2(1+Cx_+^2)^2}
\]
\begin{equation}
+\frac{BC^2[g'(x_+)(1-Cx_+^2)+x_+g''(x_+)(1+Cx_+^2)]}{(1+Cx_+^2)^2}\biggr),
\label{20}
\end{equation}
\[
g''(x_+)=\frac{2}{x_+(x_+^2+1)}.
\]
In accordance with Eq. (17) the heat capacity has a singularity when the Hawking temperature possesses an extremum ($\partial T_H(x_+)/\partial x_+=0$). Equations (16) and (17) show that at one point, $x_+=x_1$, the Hawking temperature and heat capacity become zero and the black hole remnant mass is formed. In another point $x_+=x_2$ with $\partial T_H(x_+)/\partial x_+=0$, the heat capacity has a singularity where the second-order phase transition occurs. Black holes in the range $x_2>x_+>x_1$ are locally stable but at $x_+>x_2$ black holes are unstable.
Making use of Eqs. (17), (19) and (20) the heat capacity is depicted in Fig. 3 at $C=1$.
\begin{figure}[h]
\includegraphics[height=4.0in,width=4.0in]{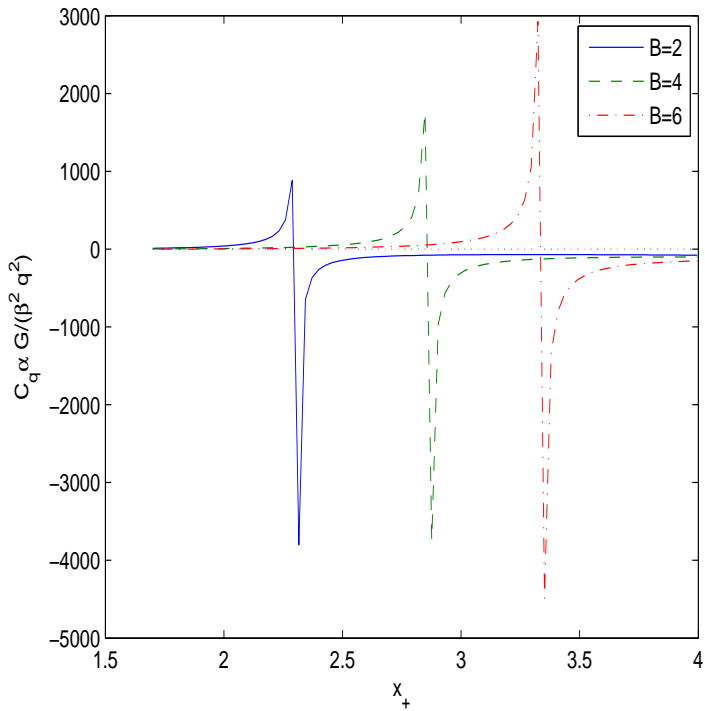}
\caption{\label{fig.3}The plot of the function $C_q(x_+)\alpha G/(\beta^2 q^2)$ at $C=1$. }
\end{figure}
The Hawking temperature and heat capacity are positive in the range $x_2>x_+>x_1$ and locally stable.

From the first law of black hole thermodynamics $dM(x_+)=T_H(x_+)dS+\phi dq$ we obtain the entropy at the constant charge \cite{Medved}
\begin{equation}
S=\int \frac{dM(x_+)}{T_H(x_+)}=\int \frac{1}{T_H(x_+)}\frac{\partial M(x_+)}{\partial x_+}dx_+.
\label{21}
\end{equation}
From Eqs. (16), (19) and (21) one finds the entropy
\begin{equation}
S=\frac{\pi(\beta q)^2}{C^2\alpha G}\int\frac{1+Cx_+^2}{x_+}dx_+=\frac{\pi r_+^2}{G}+\frac{4\pi\alpha}{G}\ln\left(\frac{r_+}{\sqrt{\beta q}}\right)+Const,
\label{22}
\end{equation}
with the integration constant $Const$. The integration constant can be chosen in the form
\begin{equation}
Const=\frac{2\pi\alpha}{G}\ln\left(\frac{\pi q\beta}{G}\right).
\label{23}
\end{equation}
Then making use of Eqs. (22) and (23) we obtain the black hole entropy
\begin{equation}
S=S_0+\frac{2\pi\alpha}{G}\ln\left(S_0\right),
\label{24}
\end{equation}
with $S_0=\pi r_+^2/G$ being the Bekenstein--Hawking entropy and with the logarithmic correction but without the coupling $\beta$. One can find same entropy (24) in other models \cite{Kruglov2,Kruglov4,Kruglov3}.

\section{Black holes shadows}

The light gravitational lensing leads to the formation of black hole shadow and a black circular disk.
The Event Horizon Telescope collaboration \cite{Akiyama} observed the image of the super-massive black hole M87*. A neutral Schwarzschild black hole shadow was studied in \cite{Synge}. We will consider photons moving in the equatorial plane, $\vartheta=\pi/2$.
With the help of the Hamilton$-$Jacobi method one obtains the equation for  the photon motion in null curves \cite{Kruglov1}
\begin{equation}
H=\frac{1}{2}g^{\mu\nu}p_\mu p_\nu=\frac{1}{2}\left(\frac{L^2}{r^2}-\frac{E^2}{f(r)}+
\frac{\dot{r}^2}{f(r)}\right)=0,
\label{25}
\end{equation}
where $p_\mu$ is the photon momentum ($\dot{r}=\partial H/\partial p_r$). The photon energy and angular momentum are constants of motion, and they are $E=-p_t$ and $L=p_\phi$, correspondingly. We can represent Eq. (25) as
\begin{equation}
V+\dot{r}^2=0, ~~~V=f(r)\left(\frac{L^2}{r^2}-\frac{E^2}{f(r)}\right).
\label{26}
\end{equation}
Photon circular orbit radius $r_p$ can be found from equation $V(r_p)=V'(r)_{|r=r_p}=0$. Making use of Eq. (26) we find
\begin{equation}
\xi\equiv\frac{L}{E}=\frac{r_p}{\sqrt{f(r_p)}},~~~f'(r_p)r_p-2f(r_p)=0,
\label{27}
\end{equation}
where $\xi$ is the impact parameter. For a distant observer as $r_0\rightarrow \infty$, the shadow radius becomes $r_s=r_p/\sqrt{f(r_p)}$ ($r_s=\xi$). By virtue of Eq. (12) and equation $f(r_+)=0$ we obtain parameters $A$, $B$ and $C$ versus $x_+$
\[
A=\frac{1+2Cx_+^2}{C^2x_+}+Bg(x_+),~~~B=\frac{AC^2x_+-2Cx_+^2-1}{C^2x_+g(x_+)},
\]
\begin{equation}
C=\frac{x_+^2+\sqrt{x_+^4+x_+(A-Bg(x_+))}}{x_+(A-Bg(x_+))},
\label{28}
\end{equation}
with $x_+=r_+/\sqrt{\beta q}$.
The functions (28) plots are depicted in Fig. 4.
\begin{figure}[h]
\includegraphics [height=4.0in,width=6.0in]{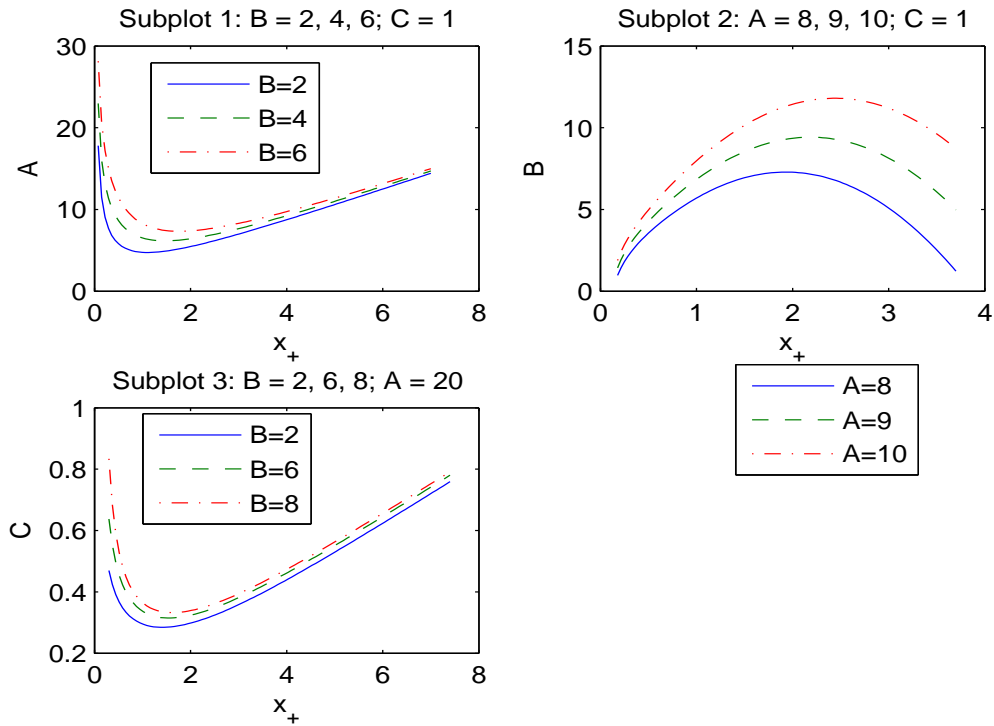}
\caption{\label{fig.4}The plots of the functions $A(x_+)$, $B(x_+)$, $C(x_+)$ }.
\end{figure}
In accordance with Fig. 4, Subplot 1, event horizon radius $x_+$ increases when parameter $A$ increases and Subplot 2 indicates that if parameter $B$ increases, the event horizon radius decreases.
According to Subplot 3 of Fig. 4, when parameter $C$ increases the event horizon radius $x_+$ also increases.

The photon sphere radii ($x_p$), the event horizon radii ($x_+$), and the shadow radii ($x_s$) for $A=15$ and $C=1$ are presented in Table 1.
It is worth noting that the null geodesics radii $x_p$ correspond to the maximum of the potential $V(r)$ ($V''\leq 0$) and belong to unstable orbits.
\begin{table}[ht]
\caption{The event horizon, photon sphere and shadow dimensionless radii for A=15, C=1}
\centering
\begin{tabular}{c c c c c c c c c c  c}\\[1ex]
\hline
$B$ & 9 & 13.5 & 14  & 15 & 16.5 & 17.5 & 18 & 19  \\[0.5ex]
\hline
 $x_+$ & 6.763 & 6.365 & 6.317 & 6.219& 6.063 & 5.953 & 5.896 & 5.777  \\[0.5ex]
\hline
 $x_p$ & 10.313 & 9.806 & 9.746 & 9.623 & 9.431 & 9.298 & 9.229 & 9.088 \\[0.5ex]
\hline
 $x_s$ & 18.311 & 17.677 & 17.603 & 17.451 & 17.216 & 17.054 & 16.971 & 16.802 \\[0.5ex]
\hline
\end{tabular}
\end{table}
Table 1 shows that when parameter $B$ increases the shadow radius $x_s$ decreases.
As $x_s>x_+$  shadow radii are defined by $r_s=x_s\sqrt{\beta q}$.

It is worth mentioning that currently there is not unique calculation of the shadow radius of M87* or SgrA* black holes within ModLogNED because our model possesses four free parameters $M$, $\alpha$, $\beta$ and $q$ (or $M$, $A$, $B$ and $C$) but from observations one knows only two values: the black hole mass and the shadow radius.

\section{Black holes energy emission rate}

The black hole shadow, for the observer at infinity, is connected with the high energy absorption cross section
\cite{Belhaj1,Wei}. At very high energies the absorption cross-section $\sigma\approx \pi r_s^2$ oscillates around the photon sphere. The energy emission rate of black holes is given by
\begin{equation}
\frac{d^2E(\omega)}{dtd\omega}=\frac{2\pi^3\omega^3r_s^2}{\exp\left(\omega/T_H(r_+)\right)-1},
\label{29}
\end{equation}
where $\omega$ is the emission frequency. By using dimensionless variable $x_+=r_+/\sqrt{\beta q}$ the black hole energy emission rate (29) becomes
\begin{equation}
\sqrt{\beta q}\frac{d^2E(\omega)}{dtd\omega}=
\frac{2\pi^3\varpi^3x_s^2}{\exp\left(\varpi/\bar{T}_H(x_+)\right)-1},
\label{30}
\end{equation}
with $\bar{T}_H(x_+)=\sqrt{\beta q}T_H(x_+)$ and $\varpi=\sqrt{\beta q}\omega$. The radiation rate versus the dimensionless emission frequency $\bar{\omega}$  for $C=1$, $A=15$ and $B=9, 14, 19$, is depicted in Fig. 5.
\begin{figure}[h]
\includegraphics[height=4.0in,width=4.0in]{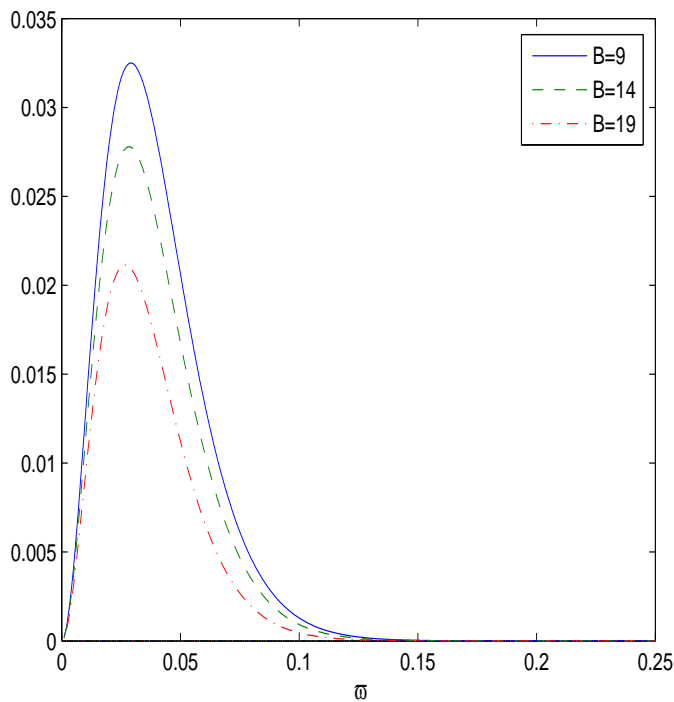}
\caption{\label{fig.5}The plot of the function $\sqrt{\beta q}\frac{d^2E(\omega)}{dtd\omega}$ vs. $\varpi$ for $B=9, 14, 19$, $A=15$, $C=1$.}
\end{figure}
Figure 5 shows that there is a peak of the black hole energy emission rate. When parameter $B$ increases, the energy emission rate peak becomes smaller and corresponds to the lower frequency. The black hole has a bigger lifetime when parameter $B$ is bigger.

\section{Quasinormal modes}

The stability of BHs under small perturbations are characterised by quasinormal modes (QNMs) with complex frequencies $\omega$. When Im~$\omega<0$ modes are stable but if Im~$\omega>0$ modes are unstable. Re~$\omega$, in the eikonal limit, is linked with the black hole radius shadow \cite{Jusufi2,Jusufi3}. Around black holes, the perturbations by scalar massless fields are described by
the effective potential barrier
\begin{equation}
V(r)=f(r)\left(\frac{f'(r)}{r}+\frac{l(l+1)}{r^2}\right),
\label{31}
\end{equation}
with $l$ being the multipole number $l=0,1,2...$. Equation (31) can be rewritten in the form
\begin{equation}
V(x)\beta q=f(x)\left(\frac{f'(x)}{x}+\frac{l(l+1)}{x^2}\right).
\label{32}
\end{equation}
Dimensionless variable $V(x)\beta q$ is depicted in Fig. 6 for $A=15$, $B=10$, $C=1$ (Subplot 1) and for $A=15$, $C=1$, $l=5$ (Subplot 2).
\begin{figure}[h]
\includegraphics{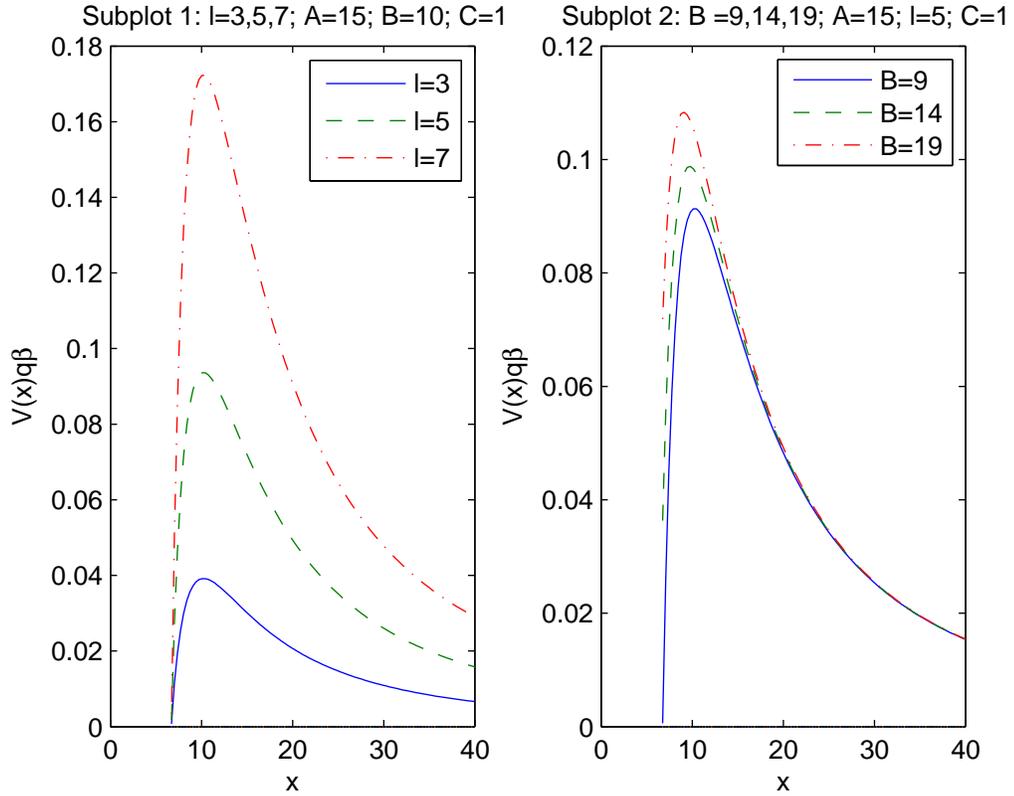}
\caption{\label{fig.6}The plot of the function $V(x)\beta q$ for $A=15$, $C=1$.}
\end{figure}
According to Figure 6, Subplot l, the potential barriers of effective potentials have maxima. For $l$ increasing the height of the potential increases. Figure 6, Subplot 2, shows that when the parameter $B$ increases the height of the potential also increases.
The quasinormal frequencies are given by \cite{Jusufi2,Jusufi3}
\begin{equation}
\mbox{Re}~\omega=\frac{l}{r_s}=\frac{l\sqrt{f(r_p)}}{r_p},~~~~\mbox{Im}~\omega=-\frac{2n+1}{2\sqrt{2}r_s}\sqrt{2f(r_p)-r_p^2f''(r_p)},
\label{33}
\end{equation}
where $r_s$ is the black hole shadow radius, $r_p$ is the black hole photon sphere radius, and $n=0,1,2,...$ is the overtone number.
The frequencies, at $A=15$, $C=1$, $n=5$, $l=10$, are given in Table 2.
\begin{table}[ht]
\caption{The real and the imaginary parts of the frequencies vs the parameter $B$ at $n=5$, $l=10$,  $A=15$, $C=1$}
\centering
\begin{tabular}{c c c c c c c c c c  c}\\[1ex]
\hline
$B$  &  14 & 15 & 16.5 & 17.5 & 18 & 19  \\[0.5ex]
\hline
 $\sqrt{\beta q}\mbox{Re}~\omega$   & 0.568 & 0.573 & 0.581 & 0.586 & 0.589 & 0.595   \\[0.5ex]
\hline
 $-\sqrt{\beta q}\mbox{Im}~\omega$  & 0.2853 & 0.2852 & 0.2849 & 0.2845 & 0.2842 & 0.2835  \\[0.5ex]
\hline
\end{tabular}
\end{table}
Because the imaginary parts of the frequencies in Table 2 are negative, modes are stable. The real part Re~$\omega$ gives the oscillations frequency. In accordance with Table 2 when parameter $B$ increasing the real part of frequency $\sqrt{\beta q}\mbox{Re}~\omega$ increases and the absolute value of the frequency imaginary part $\mid\sqrt{\beta q}\mbox{Im}~\omega\mid$ decreases. Therefore, when the parameter $B$  increases the scalar perturbations oscillate with greater frequency and decay lower.

\section{Summary}

The exact spherically symmetric solution of magnetic black holes is obtained in 4D EGB gravity coupled to ModLogNED. We studied the thermodynamics and the thermal stability of magnetically charged black holes. The Hawking temperature and the heat capacity were calculated. The phase transitions occur when the Hawking temperature has an extremum. Black holes are thermodynamically stable at some range of event horizon radii when the heat capacity and the Hawking temperature are positive. The heat capacity has a discontinuity where the second-order phase transitions take place. The black hole entropy was calculated which has the logarithmic correction. We calculated the photon sphere radii, the event horizon radii, and the shadow radii. It was shown that when the model parameter $B$ increases the black hole energy emission rate decreases and the black hole possesses a bigger lifetime. We show that when the parameter $B$ increases the scalar perturbations oscillate with greater frequency and decay lower. Other solutions in 4D EGB gravity coupled to NED were found in \cite{Kruglov2,Kruglov4,Kruglov3}.

\vspace{3mm}
\textbf{Appendix A}
\vspace{3mm}

With the spherical symmetry the energy-momentum tensor possesses the property $T_t^{~t}=T_r^{~r}$. Then, the radial pressure is $p_r=-T_r^{~r}=-\rho$. The tangential pressure $p_\perp=-T_\vartheta^{~\vartheta}=-T_\phi^{~\phi}$ is given by \cite{Dymnikova}
\[
p_\perp=-\rho-\frac{r}{2}\rho'(r),~~~~~~~~~~~~~~~~~~~~~~~~~~~~~~~~~~~~~~~~~~~~~~~~~~~~~~~~~~~~~~~~~(A1)
\]
with the prime being the derivative with respect to the radius $r$. The Weak Energy Condition (WEC) is valid when $\rho\geq 0$ and $\rho+p_k\geq 0$ (k=1,2,3) \cite{Hawking}, and then the energy density is positive. According to Eq. (7) $\rho\geq 0$. Making use of Eq. (7) we obtain
\[
\rho'(r)=-\frac{q}{\beta r^3}\ln\left(1+\frac{q\beta}{r^2}\right)-\frac{q^2}{r^3(r^2+\beta q)}\leq 0.~~~~~~~~~~~~~~~~~~~~~~~~~~~~~~~(A2)
\]
Therefore WEC, $\rho\geq 0$, $\rho+p_r\geq 0$, $\rho+p_\perp\geq 0$, is satisfied. The Dominant Energy Condition (DEC) takes place if and only if \cite{Hawking} $\rho\geq 0$, $\rho+p_k\geq 0$, $\rho-p_k\geq 0$, that includes WEC. One needs only to check the condition $\rho-p_\perp\geq 0$. By virtue of Eqs. (7), (A1) and A(2) one finds
\[
\rho-p_\perp=\frac{q}{2\beta r^2}\left[\ln\left(1+\frac{q\beta}{r^2}\right)-\frac{q\beta}{r^2+\beta q}\right].~~~~~~~~~~~~~~~~~~~~~~~~~~~~~~~~~~~~~~(A3)
\]
One can verify that $\rho-p_\perp\geq 0$ for any parameters. DEC is satisfied and therefore the sound speed is less than the speed of light. The Strong Energy Condition (SEC) is valid when $\rho+\sum_{k=1}^3 p_k\geq 0$ \cite{Hawking}. From Eqs. (8)-(10) we obtain
\[
\rho+\sum_{k=1}^3 p_k=\rho+p_\bot+p_r=p_\bot<0.~~~~~~~~~~~~~~~~~~~~~~~~~~~~~~~~~~~~~~~~~~~~~~~(A4)
\]
In accordance with Eq. (A4) SEC is not satisfied.


\begin{thebibliography}{99}

\bibitem{Clifton} T. Clifton, P. G. Ferreira, A. Padilla, and C. Skordis. Modified Gravity and Cosmology, Phys. Rept. \textbf{513}, 1 (2012) [arXiv:1106.2476].
\bibitem{Will} C. M. Will. The Confrontation between General Relativity and Experiment. Living Rev.
Rel. \textbf{17}, 4 (2014).
\bibitem{Lanczos} C. Lanczos. Elektromagnetismus als nat\"{u}rliche eigenschaft der riemannschen geometrie, Zeitschrift f\"{u}r Physik, \textbf{73},147 (1932).
\bibitem{Lanczos1}C. Lanczos. A remarkable property of the riemann-christoffel tensor in four dimensions, Annals of Mathematics, 842–850 (1938).
\bibitem{Lovelock} D. Lovelock. Divergence-free tensorial concomitants. Aequationes mathematicae, \textbf{4}, 127 (1970).
\bibitem{Lovelock1}D. Lovelock, The Einstein tensor and its generalizations, J. Math. Phys. \textbf{12}, 498 (1971).
\bibitem{Ostrogradsky}M. Ostrogradsky. M\'{e}moires sur les\'{e}quations diff\'{e}rentielles, relatives au probl\`{e}me des
isop\'{e}rim\`{e}tres. Mem. Acad. St. Petersbourg, \textbf{6}, 385 (1850).
\bibitem{Witten} D. J. Gross and E. Witten, Superstring modifications of Einstein's equations, Nucl. Phys. B \textbf{277}, 1 (1986).
\bibitem{Gross}D. J. Gross and J. H. Sloan, The quartic effective action for the heterotic string, Nucl. Phys. B \textbf{291}, 41 (1987).
\bibitem{Metsaev}R. R. Metsaev and A. A. Tseytlin, Two-loop $\beta$-function for the generalized bosonic sigma model, Phys. Lett. B \textbf{191}, 354 (1987).
\bibitem{Metsaev1}R. R. Metsaev and A. A. Tseytlin, Order $\alpha$' (two-loop) equivalence of the string equations of motion and the $\sigma$-model Weyl invariance conditions: Dependence on the dilaton and the antisymmetric tensor, Nucl. Phys. B \textbf{293}, 385 (1987).
\bibitem{Zwiebach}B. Zwiebach, Curvature squared terms and string theories, Phys. Lett. B \textbf{156}, 315 (1985).
\bibitem{Glavan} D. Glavan and C. Lin, Einstein-Gauss-Bonnet gravity in four-dimensional spacetime, Phys. Rev. Lett. \textbf{124}, 081301 (2020)
[arXiv:1905.03601].
\bibitem{Aoki} K. Aoki, M. A. Gorji, and S. Mukohyama, A consistent theory of $D\rightarrow 4$ Einstein--Gauss--Bonnet gravity Phys. Lett. B \textbf{810}, 135843 (2020) [arXiv:2005.03859].
\bibitem{Aoki1} K. Aoki, M. A. Gorji, and S. Mukohyama, Inflationary gravitational waves in consistent $D\rightarrow 4$ Einstein--Gauss--Bonnet gravity, JCAP \textbf{09}, 014 (2020) [arXiv:2005.08428].
\bibitem{Aoki2} K. Aoki, M. A. Gorji, S. Mizuno and S. Mukohyama, Inflationary gravitational waves in consistent $D\rightarrow 4$ Einstein--Gauss--Bonnet gravity, JCAP \textbf{01}, 054 (2021); JCAP  \textbf{05}, E01 (2021),  (erratum) [arXiv:2010.03973].
\bibitem{Lobo1} K. Jafarzade, M. K. Zangeneh, F. S. N. Lobo, Shadow, deflection angle and quasinormal modes of Born--Infeld charged black holes, JCAP \textbf{04}, 008 (2021) [arXiv:2010.05755].
\bibitem{Fernandes2} P. G. Fernandes, P. Carrilho, T. Clifton, and D. J. Mulryne, The 4D Einstein-–Gauss–-Bonnet theory of
gravity: a review, Class. Quant. Grav. \textbf{39}, 063001  (2022).
\bibitem{Alexeyev} S. Alexeyev, and M. Sendyuk, Black Holes and Wormholes in Extended Gravity, Universe \textbf{6}, 25 (2020).
\bibitem{Krug2} S. I. Kruglov, Magnetic black holes in AdS space with nonlinear electrodynamics, extended phase space thermodynamics and Joule-–Thomson expansion, Int. J. Geom. Meth. Mod. Phys.  \textbf{20}, 2350008 (2023) [arXiv:2210.10627].
\bibitem{Soleng}H. H. Soleng, Charged black points in General Relativity coupled to the logarithmic U(1) gauge theory,
 Phys. Rev. D \textbf{52} (1995), 6178 [arXiv:hep-th/9509033].
\bibitem{Kruglov5} S. I. Kruglov, On Generalized Logarithmic Electrodynamics, Eur. Phys. J. C \textbf{75} (2015), 88 [arXiv:1411.7741].
\bibitem{Konoplya1} R. A. Konoplya and  A. F. Zinhailo, Quasinormal modes, stability and shadows of a black hole in the 4D Einstein-Gauss-Bonnet gravity, Eur. Phys. J. C \textbf{80}, 1049 (2020) [arXiv:2003.01188].
\bibitem{Konoplya2} R. A. Konoplya and  A. F. Zinhailo, 4D Einstein--Lovelock black holes: Hierarchy of orders in curvature, Phys. Lett. B \textbf{807}, 135607 (2020).
\bibitem{Belhaj1} A. Belhaj, M. Benali, A. El Balali, H. El Moumni, and S. E. Ennadifi, Deflection Angle and Shadow Behaviors of Quintessential Black Holes in arbitrary Dimensions, Class. Quant. Grav. \textbf{37}, 215004 (2020) [arXiv:2006.01078].
\bibitem{Konoplya3} R. A. Konoplya and Z. Stuchlik, Are eikonal quasinormal modes linked to the unstable circular null geodesics, Phys. Lett. B \textbf{771}, 597 (2017) [arXiv:1705.05928].
\bibitem{Stefanov} I. Z. Stefanov, S. S. Yazadjiev, and G. G. Gyulchev, Connection between black-hole quasinormal modes and lensing in the strong deflection limit, Phys. Rev. Lett. \textbf{104}, 251103 (2010) [arXiv:1003.1609].
\bibitem{Guo1} Y. Guo and Y. G. Miao, Null geodesics, quasinormal modes and the correspondence with shadows in high-dimensional Einstein-Yang-Mills spacetimes, Phys. Rev. D \textbf{102}, 084057 (2020) [arXiv:2007.08227].
\bibitem{Wei1} S. W. Wei and Y. X. Liu, Null geodesics, quasinormal modes, and thermodynamic phase transition for charged black holes in asymptotically flat and dS spacetimes, Chin. Phys. C \textbf{44}, 115103 (2020) [arXiv:1909.11911].
\bibitem{Birkhoff} G. D. Birkhoff, \textit{Relativity and Modern Physics} (Harvard University Press, Cambrige, USA, 1923), p. 253.
\bibitem{Mignemi} S. Mignemi, N. R. Stewart, Charged black holes in effective string theory, Phys. Rev. D \textbf{47}, 5259 (1993).
\bibitem{Medved} A. J. M. Medved and E. C. Vagenas, When conceptual worlds collide: The GUP and the BH entropy,	Phys. Rev. D \textbf{70},  124021 (2004) [arXiv:hep-th/0411022].
\bibitem{Kruglov2}S. I. Kruglov, Einstein--Gauss--Bonnet gravity with nonlinear electrodynamics, Ann. Phys. \textbf{428}, 168449 (2021) [arXiv:2104.08099].
\bibitem{Kruglov4}S. I. Kruglov, Einstein–-Gauss–-Bonnet Gravity with Nonlinear Electrodynamics: Entropy, Energy Emission, Quasinormal Modes and Deflection Angle, Symmetry \textbf{13}, 944 (2021).
\bibitem{Kruglov3}S. I. Kruglov, Einstein--Gauss--Bonnet gravity with rational nonlinear electrodynamics, EPL \textbf{133}, 6 (2021) [arXiv:2106.00586].
\bibitem{Akiyama} K. Akiyama et al., First M87 Event Horizon Telescope Results, Astrophys. J.\textbf{875}, L1 (2019) [arXiv:1906.11241].
\bibitem{Synge} J. L. Synge, The escape of photons from gravitationally intense stars, Mon. Not. Roy. Astron. Soc. \textbf{131}, 463 (1966).
\bibitem{Kruglov1} S. I. Kruglov, 4D Einstein–-Gauss–-Bonnet Gravity Coupled with Nonlinear Electrodynamics, Symmetry \textbf{13}, 204 (2021).
\bibitem{Wei} S. W. Wei and Y. X. Liu, Observing the shadow of Einstein-Maxwell-Dilaton-Axion black hole, JCAP 11, 063 (2013) [arXiv:1311.4251].
\bibitem{Jusufi2} K. Jusufi, Quasinormal Modes of Black Holes Surrounded by Dark Matter and Their Connection with the Shadow Radius, Phys. Rev. D \textbf{101}, 084055 (2020) [arXiv:1912.13320].
\bibitem{Jusufi3} K. Jusufi, Connection Between the Shadow Radius and Quasinormal Modes in Rotating Spacetimes, Phys. Rev. D \textbf{101}, 124063 (2020) [arXiv:2004.04664].
\bibitem{Dymnikova} I. Dymnikova, Regular electrically charged vacuum structures with de Sitter centre in nonlinear electrodynamics coupled to general relativity, Class. Quant. Grav. \textbf{21}, 4417 (2004) [arXiv:gr-qc/0407072].
\bibitem{Hawking} S. W. Hawking and G. F. R. Ellis, \textit{The large scale structure of space-time}, Cambridge Univ. Press, Cambridge UK (1973).

\end{thebibliography}
\end{document}